# Observation of magnetism induced topological edge state in antiferromagnetic topological insulator MnBi$_4$Te$_7$

*Hao-Ke Xu[1], Mingqiang Gu[2], Fucong Fei[3], Yi-Sheng Gu[1], Dang Liu[1], Qiao-Yan Yu[1], Sha-Sha Xue[1], Xu-Hui Ning[1,4], Bo Chen[3], Hangkai Xie[3], Zhen Zhu[1], Dandan Guan[1], Shiyong Wang[1], Yaoyi Li[1], Canhua Liu[1], Qihang Liu[2], Fengqi Song[3], Hao Zheng[1*] and Jinfeng Jia[1*]*

1. School of Physics and Astronomy, Key Laboratory of Artificial Structures and Quantum Control (Ministry of Education), Shenyang National Laboratory for Materials Science, Tsung-Dao Lee Institute, Shanghai Jiao Tong University, Shanghai 200240, China

2. Shenzhen Institute for Quantum Science and Engineering and Department of Physics, Southern University of Science and Technology, Shenzhen 518055, China

3. National Laboratory of Solid State Microstructures, Collaborative Innovation Center of Advanced Microstructures, and College of Physics, Nanjing University, Nanjing 210093, China.

4. Zhiyuan College, Shanghai Jiao Tong University, Shanghai 200240, China

**ABSTRACT**:

Breaking time reversal symmetry in a topological insulator may lead to quantum anomalous Hall effect and axion insulator phase. MnBi$_4$Te$_7$ is a recently discovered antiferromagnetic topological insulator with $T_N$ ~12.5 K, which is constituted of alternatively stacked magnetic layer (MnBi$_2$Te$_4$) and non-magnetic layer (Bi$_2$Te$_3$). By means of scanning tunneling spectroscopy, we clearly observe the electronic state present at a step edge of a magnetic MnBi$_2$Te$_4$ layer but absent at non-magnetic Bi$_2$Te$_3$ layers at 4.5 K. Furthermore, we find that as the temperature rises above $T_N$, the edge state vanishes, while the point defect induced state persists upon temperature increasing. These results confirm the observation of magnetism induced edge states. Our analysis based on an axion insulator theory reveals that the nontrivial topological nature of the observed edge state.

H.-K. Xu, M. Gu and F. Fei contributed equally to this work.

\* haozheng1@sjtu.edu.cn; jfjia@sjtu.edu.cn

**MAIN TEXT:**

The key characteristics of topological non-trivial materials is bulk-boundary correspondence. For example, in a two-dimensional (2D) quantum spin hall insulator, the strong spin-orbit coupling inverts the conventional band order between the conduction band and the valence band and opens a nontrivial band gap in bulk electronic structure, leading to a gapless state at the material's boundary [1-4]. Scanning tunneling microscopy/spectroscopy (STM/S) plays a key role in topological matter study[5-8], especially has been employed to directly visualize the topological edge states. To-date, most of these measurements are performed in quantum spin hall states, for example, Bi, $WTe_2$, $ZrTe_5$ and others [9-17]. Many theoretically predicted unique phenomena of the quantum spin hall edge state, such as the back-scattering forbidden effect, helical Tomonaga–Luttinger liquid state, were also discovered using STM and STS [10, 12].

The axion insulator is an important class of topological nontrivial material, characterized by the effective Chern-Simons term with a quantized bulk magnetoelectric coupling coefficient. Once the magnetism opens a gap at the boundary of the axion insulator, the Chern-Simons term ensures a non-trivial phase with half-quantized anomalous Hall effect [18]. Unlike the study of 2D quantum spin Hall insulator, direct demonstration of such bulk-boundary correspondence in an axion insulator still remains elusive. Zero Hall conductance and resistance plateaus have been suggested as evidence to prove the axion insulator state in a magnetically doped topological insulator heterostructure [19,20]. However, debates have been raised that such transport measurements are not sufficient to distinguish the axion insulator phase from the trivial insulator phase [21-23], half-quantized anomalous hall effect, which supposed as a fingerprint of axion, is still unrevealed [24,25]. Recently, $MnBi_{2n}Te_{3n+1}$ class of stoichiometry materials have been theoretically predicted and experimentally proven as time-reversal symmetry breaking topological insulators [26-52]. Various topological phases are predicted or measured in such material family, including the axion insulator state [27-31,53-58]. Detection of the bulk-boundary correspondence in this series of materials using STM/S is a valuable task.

In this work, we apply temperature-dependent STM/S to study the magnetic and nonmagnetic layers of $MnBi_4Te_7$ and clearly find the magnetism-induced electronic state on magnetic layer's edge. Our model analysis certifies the nontrivial Chern number of our observed topological edge state.

**RESULTS AND DISCUSSION**

$MnBi_4Te_7$ is a layered material. As shown in Fig. 1, $Bi_2Te_3$ and $MnBi_2Te_4$ layers are alternatively stacked along c axis. ($MnBi_2Te_4$ and $Bi_2Te_3$ layers are denoted as 124 and 023 respectively, in the following.) The sample is easily cleaved due to the weak Van Der Waals interaction between layers. After cleavage, both 124 and 023 terrace can be found in our experiment. The step heights of the 124 and 023 layers are 1.3 nm and 1.0 nm, respectively. The spin of the Mn atoms aligns ferromagnetically within each 124 layer, while antiferromagnetically between adjacent 124 layers. In other words, an A-type antiferromagnetic (AFM) order is formed. Indeed, transport measurements confirm the antiferromagnetism in our samples and demonstrate that the Néel temperature is around 12.5 K (Fig. S1). On the STM images (Figs. 1b and c), we reveal that both 124 and 023 surfaces are clean and atomically flat, but feature different types of point defects. For example, dark triangle shaped defects can be found on both 124 and 023 surfaces, but bright dotted shaped defects can only be seen on a 124 surface. According to a previous STM research [46], we attribute the former to the Te atom on the top surface being replaced by Bi, and the later to Mn-Bi anti-site.

The electronic characteristics on both surfaces are then evaluated using dI/dV spectra. Our spectral shapes on both surfaces are consistent with recent reports [45,46]. In Fig. 1d, we observe a "V" shaped feature in the energy range of 0.2 eV to -0.4 eV, with the bottom of the V at -0.27 eV. To further explain our findings, we use first-principles calculations to simulate the surface band structure on 124 and 023 surfaces. We find that a single Dirac cone-shaped surface state with its Dirac node at -0.27 eV according to the minimum of the local density of state (LDOS), *i.e.* the bottom of the V. The valence band maximum and conduction band minimum in our sample are at -0.14 eV and -0.36 eV respectively, based on the comparison of our dI/dV spectra and band structure maps in Figs.1 and S2. On both 124 and 023 surfaces, pure topological surface states without mixture of bulk bands can be identified within this energy range.

Step edge is a boundary of a material and consequently an indicator of the bulk topology in a crystal. We first pay attention to a step with 1.3 nm height, indicating that the upper and lower terraces belong to a 124 and a 023 surface respectively. We measure the position-dependent dI/dV spectra, which traverses a 1.3 nm height step and is 20 nm long (Fig. 2). The LDOS at the step edge drastically differs from the one within the terrace, indicating the existence of an edge state. When the spectrum is moved away from the edge, such character vanishes. Moreover, the peak positions, *i. e.* -0.27 eV, corresponds to the energy level of the surface Dirac node. We suggest it is the topological edge state in $MnBi_4Te_7$. More evidences shall be offered in the following.

As a controlled experiment, we repeat the position-dependent dI/dV measurement on a 1.0 nm high step that separates an upper non-magnetic 023 terrace and a lower 124 terrace. Our results in Figs. 3 a and b reveal that the dI/dV spectra taken at the step edge have almost same shapes as the ones measured within terrace. In another word, we don't find evidences of edge states on a 023 layer.

All of the above experiments are carried out at 4.5 K, which is lower than our $MnBi_4T_7$ sample's Néel temperature of 12.5 K. We find that the electronic edge state is only present on the magnetic 124 terrace but not on the nonmagnetic terrace. To better understand the relationship between the edge state and magnetism, we now raise the system temperature to 77 K, driving the MnBi$_4$T$_7$ crystal into a nonmagnetic (paramagnetic) state. The LDOS obtained on and off step edge (Fig. 3 c and d) have nearly identical shapes (especially near the energy of surface Dirac node). Despite the fact that data quality at 77 K degrades due to the increased thermal fluctuation, our results still can conclude that the edge state on a 124 terrace vanishes at this temperature.

MnBi$_4$Te$_7$ is a complex material, which possesses many defects. Previous works discovered that a specific sort of defect in MnBi$_4$Te$_7$ can also produce a Dirac resonance state [46,59]. We successfully repeat the prior result in Fig. 4. However, we clearly discern that the Dirac resonance defect state, on the other hand, does not change at 77 K, which contrasts sharply with the temperature dependent evolution of the step edge state.

We note that during cleaving, the sample breaks at the step edge resulting in numerous defects. For example, we observe lots of bright dots on the step edge in Fig. 2b, which can be attributed to point defects. The defects may induce extra electronic states, such as additionally peaks in LDOS.

In order to clarify the influence of the defect states, we conduct an in-depth investigation into the edge states. We focus on a step of a 124 layer (Fig. 5), and intentionally take a number of dI/dV spectra at many positions along the step edge. In Fig. 5b, we observe that the dI/dV spectra take various shapes. Moreover, the LDOS peaks are not always located at -0.27 eV, which is the energy of the Dirac node measured on a defect free region. In principle, a topological edge state should appear at the Dirac energy. But in reality, some sort of defects may be able to change the local potential and thus shift the Dirac node energy away from the value -0.27 eV (by several tens of meV). We emphasize that all of our spectra have a peak close to (although not exactly at) -0.27 eV. More importantly, we raise up the temperature to 77 K, and record again a series of dI/dV spectra along a 124 magnetic layer edge (Fig. 5 c and d). In this situation, one can find no LDOS peak close to the Dirac node energy (around -0.27 eV). In contrast, certain peaks located between -0.1 eV to -0.2 eV remain, which can be attributed different types of defect states. Based on the fact that the LDOS peaks close to -0.27 eV appear everywhere on a 124 step edge at 4.5 K but disappear at 77 K, we believe that their origins are the topological edge state.

We summarize our experimental results, *i. e.* the dI/dV spectra and maps on distinct terminations of our MnBi$_4$Te$_7$ sample under varied measurement temperatures in Fig. 6 and conclude that edge state appears at the boundary of magnetic layer only below MnBi$_4$Te$_7$ 's Néel temperature.

Theoretically speaking, bulk-boundary correspondence is the key feature of various topological phases, and it takes on distinct forms in each one. There are two mechanisms that support a topological boundary mode in an axion insulator: (i) a hinge between the top surface and the side surface. Both surfaces of an axion insulator are gapped, resulting in a local Chern marker $C_{top} = \pm \frac{1}{2}$ and $C_{side} = \mp \frac{1}{2}$. In this sense, the hinge carries a quantized anomalous Hall conductivity, *i. e.* $\sigma_{edge} = (C_{top} - C_{side})\frac{e^2}{h} = \pm \frac{e^2}{h}$; (ii) a surface magnetic domain wall, as shown in Figs. 6 and S3, that divides two different magnetic domains. When a domain wall is crossed, the spin orientation is flipped, leading to a chiral mode manifesting as the changes in the Chern marker, *i. e.* $\Delta C = \frac{1}{2} - \left(-\frac{1}{2}\right) = 1$ in an axion insulator[60]. To depict this domain wall, an effective Hamiltonian $H(x,y) = -iv(\partial_y \sigma_x - \partial_x \sigma_y) - \text{sgn}(y)M_z\sigma_z$ can be employed, leading to a cross-gap boundary state located at the domain wall. Our observed magnetism-induced step edge state is consistent with of the axion insulator state.

We note that two very recent works have reported the observation of edge states in thin film $MnBi_2Te_4$, with the thickness ranging from 1 to 6 layers [61,62]. The underlying mechanism of these works can be attributed case (i). However, we believe that the origin of the observed edge states in our antiferromagnetic $MnBi_4Te_7$ is related to mechanism (ii) for two reasons. First, because of the $T\tau_{1/2}$-symmetry, the center of the side surface in $MnBi_4Te_7$ possess a gapless Dirac cone. As a result, the gap closing process from the center of top surface to the center of the side surface is progressive, with no abrupt boundary [31]. Second, the edge states can only be observed on a 124 terrace edge but not on 023 terrace. The 124 step edge is magnetic domain wall as well as a geometrical step edge, whereas the 023 step edge is not. Furthermore, temperature dependent measurement validates the magnetic origin of the edge states. Taken these evidences together, we believe that the surface magnetic domain wall, which emerges naturally and inevitably at the magnetic layers edge, is a critical ingredient to our discovery.

**CONCLUSION**

In a summary, our symmetric termination- and temperature- dependent STM/S measurements, together with DFT simulation and model analysis, proves that the magnetism domain wall induced topological edge state exists in a $MnBi_4Te_7$ sample.

**METHODS:**

**Sample growth**

$MnBi_4Te_7$ single crystals were grown by the flux method. The raw materials with the molar ratio of MnTe: $Bi_2Te_3$ = 1:7.2 were placed into an alumina crucible and vacuum-sealed by a quartz ampoule. The ampoule was then heated up to 1000 °C in 48 hours by a muffle furnace and maintains at this temperature for 48 hours. After a slowly cool-down process to 580 °C in 72 hours, the as-grown crystals were separated from the excess $Bi_2Te_3$ flux by centrifugation.

**STM measurement**

The STM measurements were performed in a commercial STM (Unisoku 1600) operating in ultra-high vacuum circumstance, whose base pressure was maintained around $3 \times 10^{-10}$ mBar. Operating temperature was maintained approximately at 4.5 K using liquid helium as refrigerant or at 77 K using liquid nitrogen. The $MnBi_4Te_7$ sample we measured was cleaved *in situ* at room temperature. STM tip was produced by electrochemically etched Tungsten wire, which was annealed by electron beam heating. The tunnelling differential conductance (dI/dV) signal was obtained by standard Lock-in technique, where 5 mV, 991 Hz were chosen as bias amplitude and oscillation frequency.

**Theoretical study**

DFT calculations were performed using the projector-augmented wave [63] pseudopotentials within the scheme of Perdew-Burke-Ernzerhof (PBE) [64] form of GGA approach, as implemented in the Vienna ab-initio Simulation Package (VASP) [65,66]. PBE+U method [67] with U(Mn) = 5 eV was used to consider the correlation effect of the Mn-d electrons. The Brillouin zone was sampled by a 6×6×3 Γ-centered Monkhorst-Pack k-point mesh. The energy cutoff and total energy tolerance for the self-consistent calculations were 500 eV and $10^{-6}$ eV, respectively. The lattice constants ($a_0$ = 4.355 Å and $c_0$ = 47.63 Å) were fixed in all the calculations, and the atomic positions were fully relaxed until the force on each atom is less than $10^{-1}$ eV/Å. Spin-orbit coupling was included in the calculations self-consistently. The spectral functions for various surface terminations were computed using iterative Green's function method as implemented in the Wannier Tools package [68]. Tight-binding models within the Wannier representations by projecting the Bloch states onto Mn-d, Bi-p, and Te-s orbitals were constructed with the WANNIER90 code interfaced to VASP [69, 70].

**ACKNOWLEDGEMENTS:**

We thank the NSFC (Grants No. 11790313, No. 92065201, No. 11874256, No. 11874258, No. 12074247, No. 12174252, No. 11861161003, No. 12025404, and No. 11904165), Ministry of Science and Technology of China (Grants No. 2019YFA0308600, 2020YFA0309000, No. 2017YFA0303203), the Strategic Priority Research Program of Chinese Academy of Sciences (Grant No. XDB28000000) and the Science and Technology Commission of Shanghai Municipality (Grants No. 2019SHZDZX01, No. 19JC1412701, No. 20QA1405100), the Natural Science Foundation of Jiangsu Province (No. BK20190286) for partial support.

**SUPPORTING INFORMATION:**

Supporting Information available at

Fig. S1: resistivity and magnetization of MnBi$_4$Te$_7$; Fig. S2: electronic structure of bulk and two termination surface of MnBi$_4$Te$_7$; Fig. S3: schematic image shows different spin-flop behaviors for two step edges of MnBi$_4$Te$_7$ (PDF).

FIGURES:

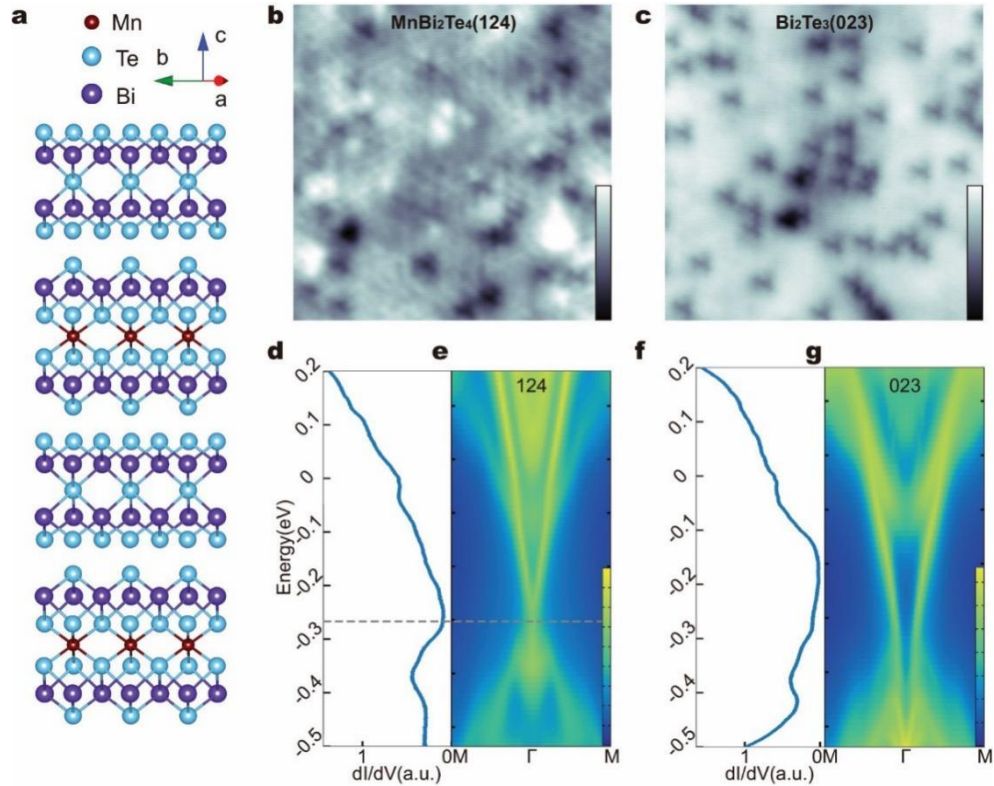

**Figure 1: Surface topography and band structure of the antiferromagnetic topological insulator MnBi$_4$Te$_7$.** **a**, Crystal structure of two MnBi$_4$Te$_7$ unit-cells, which is stacked by MnBi$_2$Te$_4$ and Bi$_2$Te$_3$ layers. **b** and **c**, STM images representing the typical morphologies of MnBi$_2$Te$_4$(124) surface and Bi$_2$Te$_3$(023) surface respectively. (500mV, 200pA). **d** and **e**, averaged dI/dV spectra from 30 single dI/dV spectra obtained on different locations (200mV, 300pA) and simulation of surface spectral weight on 124 surface, respectively. Grey dashed line indicates the energy of surface Dirac node. **f** and **g** are same as d and e but on 023 surface.

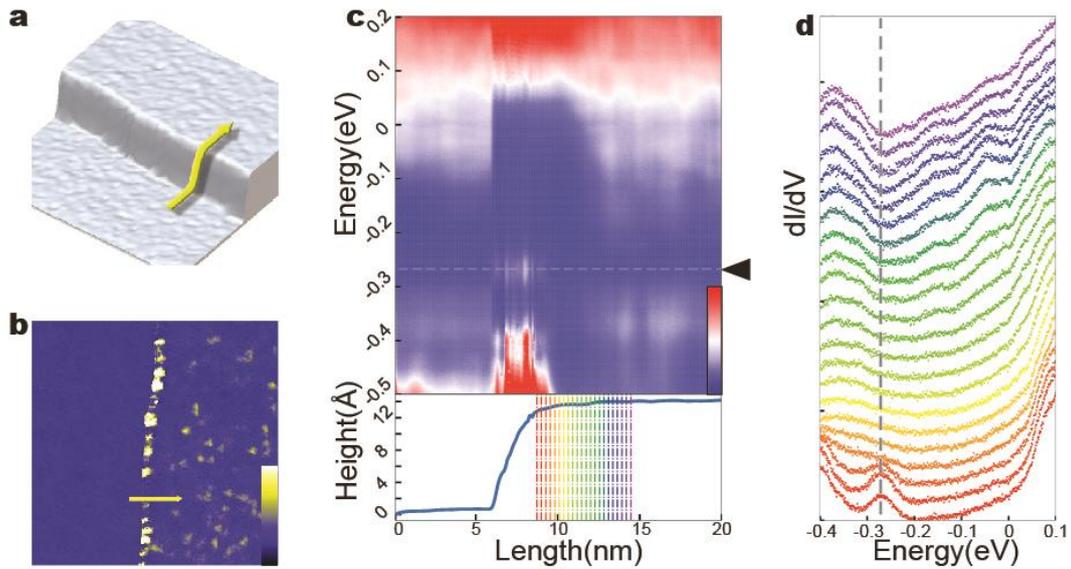

**Figure 2: Edge state on 124 surface measured at liquid Helium temperature. a**, 3D representation of an STM image on a MnBi$_4$Te$_7$ with a 1.3nm heigh step (100 × 100 nm$^2$, 200mV, 300pA). **b**, dI/dV map on the same area as the image in a at the energy of Dirac node. **c**, Diagram of position-dependence dI/dV spectra (set points: 200mV, 300pA) obtained along the yellow line in a and b. Corresponding topographic line-profiler is shown in the bottom panel. Edge state lays at the energy of Dirac node, which is marked with grey dashed line. **d**, 22 spectra extracted from c with the spatial resolution better than 0.3nm. The spectra displayed from bottom to top are obtained starting from the edge toward the terrace, whose positions are indicated by the rainbow lines in c. Dashed line guides the eyes to the edge state.

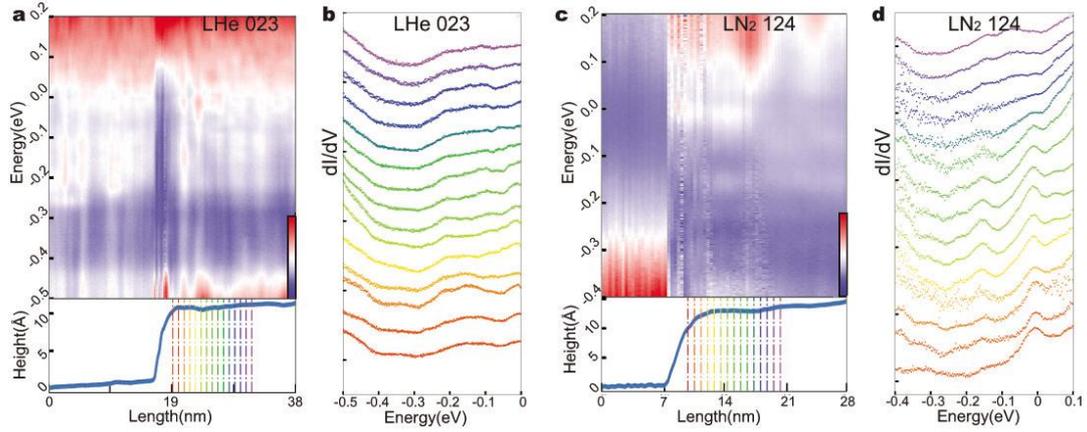

**Figure 3: Absence of edge state on 023 surface and disappearance of edge state on 124 surface at liquid Nitrogen temperature. a**, diagram of the position-dependence dI/dV spectra measured by crossing step with 1.0nm height at 4.5K.(set points :200mV, 300pA ). **b**, a series of spectra taken out of a, acquired by starting from step edge toward terrace. Positions of the spectra are indicated by the rainbow dashed lines in a. No obvious edge state can be detected. **c** and **d** are similar as a and b, but were measured across step with 1.3nm height when the system was heated up to 77K (above the Neel temperature of MnBi$_4$Te$_7$). The edge state which can be observed at 4.5K now disappears.

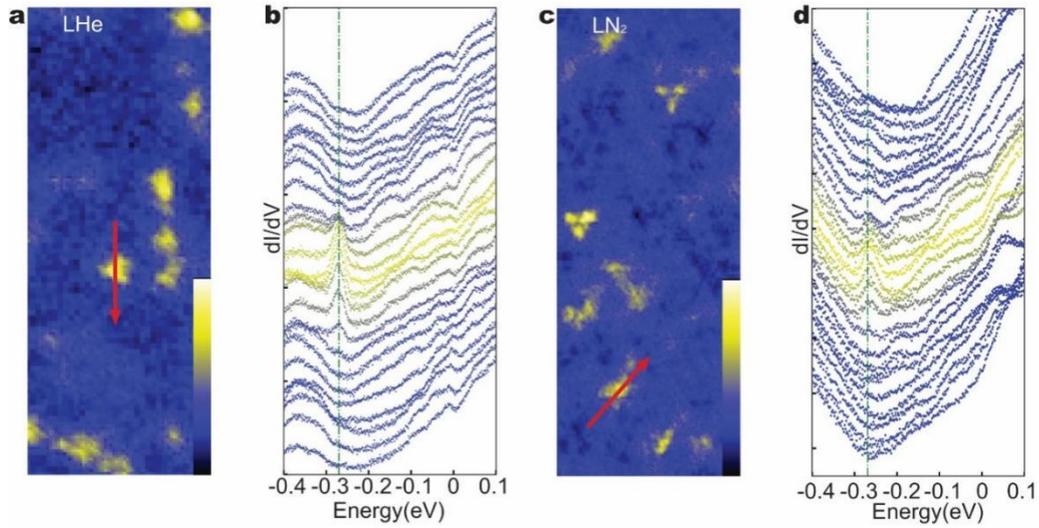

**Figure 4: Persistence of the defect-induced state upon temperature increasing. a**, a dI/dV map ( 50 × 19.5nm$^2$, 270mV, 300pA) displaying several defects on a 124 terrace at 4.5K. **b**, a set of dI/dV spectra collected around a defect, *i. e.* along the red line in a. LDOS peaks appear at the energy of Dirac node on top of the defect. **c**, a dI/dV map ( 50 × 19.5nm$^2$, 270mV, 100pA) measuring a 124 terrace but at 77K. **d**, a series of dI/dV spectra taken around the defect in c. The defect state, in contrast to the loss of the edge state, stays at 77K.

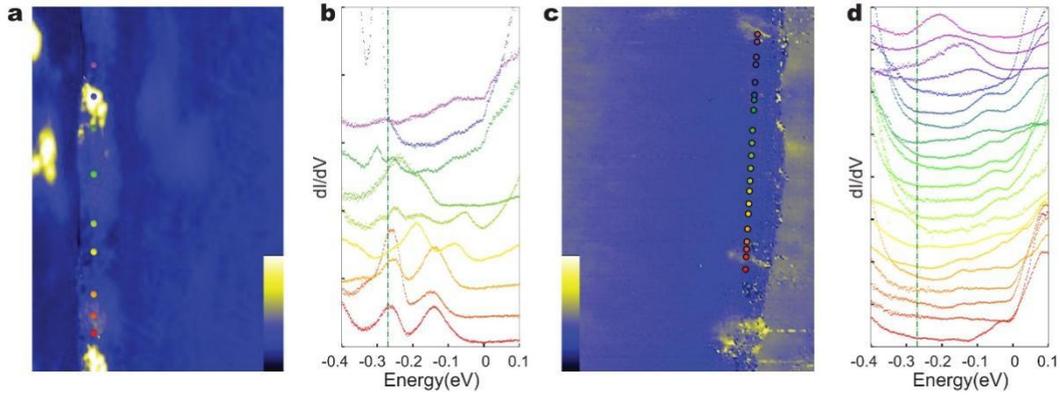

**Figure 5: Confirmation of the edge state against defect state. a**, a dI/dV map (13.3 ×20nm², -0.3V 200pA) on a 1.3nm high step at 4.5K. The right part is a 124 terrace. **b**, a set of dI/dV spectra taken at the colored spots in a. Dotted lines in b indicate the Dirac node energy. It's worth noting that although the defects at the step edge may change the local environment, all spectra show enhanced dI/dV intensities (peak or hump) close to the Dirac energy. It proves the presence of an edge state at each position. **c**, a dI/dV map (20×30nm², -0.27V 200pA) on a 1.3nm high step but at 77K. The left part is a 124 terrace. **d**, dI/dV spectra measured on c, whose positions are indicated by the colored dots. On can reveal that the defect states remain in some locations, but the edge state disappears everywhere.

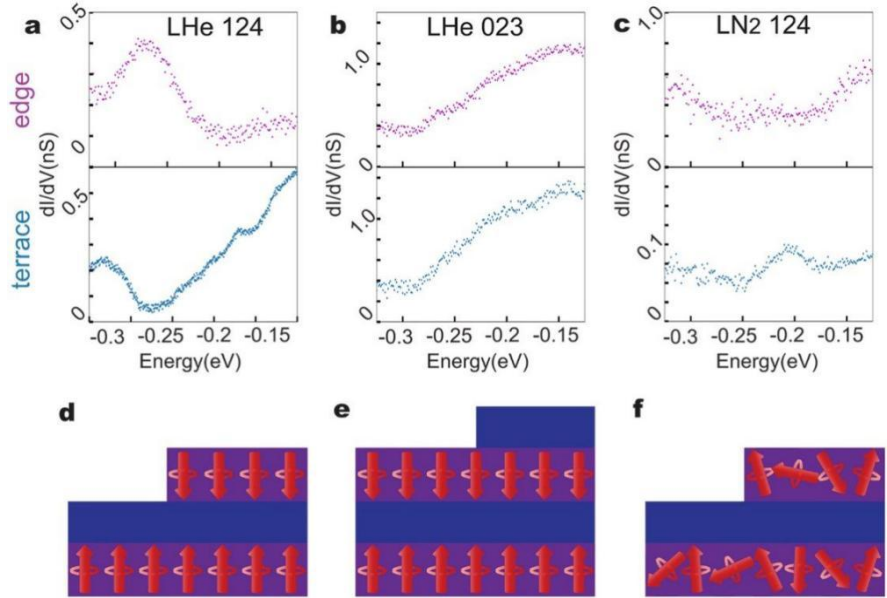

**Figure 6: Magnetic domain wall induced topological edge state in the Axion insulator MnBi$_4$Te$_7$. a - c**, dI/dV spectra exhibiting the local density of state at the step edge (upper panel) and terrace (lower panel) measured on 124 surface at 4.5K, on 023 surface at 4.5K and on 124 surface at 77K respectively. The electronic edge state only appears in a. **d - f**, Schematic models of three kinds of steps corresponding to a, b and c. Purple (blue) rectangles stand for 124 (023) layers. Red arrows indicate the spins in Mn atoms. d and e are in antiferromagnetic order, while f is in paramagnetic states. We note that only the step in d serves as both a step edge and a magnetic domain wall.

**References:**

1. Kane, C. L.; & Mele, E. L. Quantum Spin Hall Effect in Graphene. *Phys. Rev. Lett.* **2005**, 95, 226801.

2. Bernevig, B. A. & Zhang, S. C. Quantum Spin Hall Effect. *Phys. Rev. Lett.* **2006**, 96, 106802.

3. König, M.; Wiedmann, S.; Brüne, C.; Roth, A.; Buhmann, H.; Molenkamp, L. W.; Qi, X. L.; Zhang, S. C. Quantum Spin Hall Insulator State in HgTe Quantum Wells. *Science* **2007**, 318, 766-770.

4. Wu, S.; Fatemi, V.; Gibson, Q. D.; Watanabe, K.; Taniguchi, T.; Cava, R. J.; Jarillo-Herrero, P. Observation of the Quantum Spin Hall Effect up to 100 Kelvin in a Monolayer Crystal. *Science* **2018**, 359, 76-79.

5. Zheng, H.; Xu, S. Y.; Bian, G.; Guo, C.; Chang, G. Q.; Sanchez, D. S.; Belopolski, I.; Lee, C. C; Huang, S. M.; Zhang, X. et al. Atomic-Scale Visualization of Quantum Interference on a Weyl Semimetal Surface by Scanning Tunneling Microscopy. *ACS Nano* **2016**, 10, 1378–1385.

6. Zhu, Z.; Chang, T. R.; Huang, C. Y.; Pan, H. Y.; Nie, X. A.; Wang, X. Z.; Jin, Z. T.; Xu, S. Y.; Huang, S. M.; Guan, D. D.; Wang, S. Y.; Li, Y. Y.; Liu, C. H.; Qian, D.; Ku, W.; Song, F. Q.; Lin, H.; Zheng, H. & Jia, J. F. Quasiparticle Interference and Nonsymmorphic Effect on a Floating Band Surface State of ZrSiSe. *Nat Commun* **2018**, 9, 4153.

7. Nie, X. A.; Li, S. J.; Yang, M.; Zhu, Z.; Xu, H. K.; Yang, X.; Zheng, F. W.; Guan, D. D.; Wang, S. Y.; Li, Y. Y.; Liu, C. H.; Li, J.; Zhang, P.; Shi, Y. G.; Zheng, H. & Jia, J. F. Robust Hot Electron and Multiple Topological Insulator States in $PtBi_2$.

*ACS Nano* **2020**, 14, 2366-2372.

8. Zhu, Z.; Papaj, M.; Nie, X. A.; Xu, H. K.; Gu, Y. S.; Yang, X.; Guan, D. D.; Wang, S. Y.; Li, Y. Y.; Liu, C. H.; Luo, J. L.; Xu, Z. A.; Zheng, H.; Fu, L. & Jia, J. F. Discovery of Segmented Fermi Surface Induced by Cooper Pair Momentum. *Science* **2021**, 374, 1381-1385.

9. Yang, F.; Wang, Z. F.; Yao, M. Y.; Zhu, F. F.; Song, Y. R.; Wang, M. X.; Xu, J. P.; Fedorov, A. V.; Sun, Z.; Zhang, G. B.; Liu, C. H.; Qian, D.; Gao, C. L. & Jia, J. F. Spatial and Energy Distribution of Topological Edge States in Single Bi(111) Bilayer. *Phys. Rev. Lett.* **2012**, 109, 016801.

10. Drozdov, I. K.; Alexandradinata, A.; Jeon, S. J.; Perge, S. N.; Ji, H. W.; Cava, R. J.; Bernevig, B. A. & Yazdani, A. One-Dimensional Topological Edge States of Bismuth Bilayers. *Nat. Phys.* **2014**, 10, 664-669.

11. Reis, F.; Li, G.; Dudy, L.; Bauernfeind, M.; Glass, S.; Hanke, W.; Thomale, R.; Schäfer, J. & Claessen, R. Bismuthene on a SiC Substrate: A Candidate for a High-Temperature Quantum Spin Hall Material. *Science* **2017**, 357, 287-290.

12. Stühler, R.; Reis, F.; Müller, T.; Helbig, T.; Schwemmer, T.; Thomale, R.; Schäfer, J. & Claessen, R. Tomonaga–Luttinger Liquid in the Edge Channels of a Quantum Spin Hall Insulator. *Nat. Phys.* **2020**, 16, 47-51.

13. Peng, L.; Yuan, Y.; Li, G.; Yang, X.; Xian, J. J.; Yi, C. J.; Shi, Y. G. & Fu, Y. S. Observation of Topological States Residing at Step Edges of $WTe_2$. *Nat. Commun.* **2017**, 8, 659.

14. Tang, S. J.; Zhang, C. F.; Wong, D.; Pedramrazi, Z.; Tsai, H. Z.; Jia, C. J.; Moritz,

B.; Classen, M.; Ryu, H.; Kahn, S.; Jiang, J.; Yan, H.; Hashimoto, M.; Lu, D. H.; Moore, R. G.; Hwang, C. C.; Hwang, C. .C; Hussain, Z.; Chen, Y. L.; Ugeda, M. M. et al. Quantum Spin Hall State in Monolayer 1T'-WTe$_2$. *Nat Phys.* **2017**, 13, 683-687.

15. Sessi, P.; Sante, D. D.; Szczerbakow, A.; Glott, F.; Wilfert, S.; Schmidt, H.; Bathon, T.; Dziawa, P.; Greiter, M.; Neupert, T.; Sangiovanni, G.; Story, T.; Thomale, R. & Bode, M. Robust Spin-Polarized Midgap States at Step Edges of Topological Crystalline Insulators. *Science* **2016**, 354, 1269-1273.

16. Li, X. B.; Huang, W. K.; Lv, Y. Y.; Zhang, K. W.; Yang, C. L.; Zhang, B. B.; Chen, Y. B.; Yao, S. B.; Zhou, J.; Lu, M. H.; Sheng, L.; Li, S. C.; Jia, J. F.; Xue, Q. K.; Chen, Y. F. & Xing, D. Y. Experimental Observation of Topological Edge States at the Surface Step Edge of the Topological Insulator ZrTe$_5$. *Phys. Rev. Lett.* **2016**, 116, 176803.

17. Wu, R.; Mam, J. Z.; Nie, S. M.; Zhan, L. X.; Huang, X.; Yin, J. X.; Fu, B. B.; Ricahrd, P.; Chen, G. F.; Fang, Z.; Dai, X.; Weng, H. M.; Qian, T.; Ding, H. & Pan, S. H. Evidence for Topological Edge States in a Large Energy Gap near the Step Edges on the Surface of ZrTe$_5$. *Phys. Rev. X* **2016**, 6, 021017.

18. Armitage, N. P. & Wu, L. On the Matter of Topological Insulators as Magnetoelectrics. *SciPost Phys.* **2019**, 6, 046

19. Mogi, M.; Kawamura, M.; Tsukazaki, A.; Yoshimi, R.; Takahashi, K. S.; Kawasaki, M.; Tokura, Y. Tailoring Tricolor Structure of Magnetic Topological Insulator for Robust Axion Insulator. *Sci. Adv.* **2017**, 3, eaao1669.

20. Xiao, D.; Jiang, J.; Shin, J. H.; Wang, W.; Wang, F.; Zhao, Y. F.; Liu, C.; Wu, W.; Chan, M. H. W.; Samarth, N.; Chang, C. Z. Realization of the Axion Insulator State in Quantum Anomalous Hall Sandwich Heterostructures. *Phy. Rev. Lett.* **2018**, 120, 056801.

21. Fijalkowski, K. M.; Liu, N.; Hartl, M.; Winnerlein, M.; Mandal, P.; Coschizza, A.; Fothergill, A.; Grauer, S.; Schreyeck, S.; Burnner, K.; Greiter, M.; Thomale, R.; Gould, C. & Molenkamp, L. W. Any Axion Insulator must be a Bulk Three-Dimensional Topological Insulator. *Phys. Rev. B* **2011**, 103, 235111.

22. Kou, X. F.; Pan, L.; Wang, J.; Fan, Y. B.; Choi, E. U.; Lee, W. L.; Nie, T. X.; Murata, K.; Shao, Q. M.; Zhang, S. C. & Wang, K. L. Metal-to-Insulator Switching in Quantum Anomalous Hall States. *Nat. Commun.* **2015**, 6, 8474.

23. Zhao. Y. F. & Liu, Q. H.; Routes to Realize the Axion-Insulator Phase in MnBi$_2$Te$_4$(Bi$_2$Te$_3$)n Family. *Appl. Phys. Lett.* **2021**, 119, 060502.

24. Qi, X. L.; Hughes, T. L. & Zhang, S. C. Topological Field Theory of Time-Reversal Invariant Insulators. *Phys. Rev. B* **2008**, 78, 195424.

25. Essin A. M.; Moore, J. E. & Vanderbilt D. Magnetoelectric Polarizability and Axion Electrodynamics in Crystalline Insulators. *Phys. Rev. Lett.* **2009**, 102, 146805.

26. Li, J .H.; Li, Y.; Du, S. Q.; Wang, Z,. Gu, B. L.; Zhang, S. C.; K. He.; Duan, W. H.; Xu, Y. Intrinsic Magnetic Topological Insulators in Van Der Waals Layered MnBi$_2$Te$_4$-Family Materials. *Sci. Adv* **2019**, 5, eaaw5685.

27. Otrokov, M. M.; Klimovskikh, I. I.; Bentmann, H.; Estyunin, D.; Zeugner, A.;

Aliev, Z. S.; Gaß, S.; Wolter, A. U. B.; Vyazovskaya, A. Y.; Eremeev, S. V.; Koroteev, Y. M.; Kuznetsov, V. M.; Freyse, F.; Sánchez-Barriga, J.; Amiraslanov, I. R.; Babanly, M. B.; Mamedov, N. T.; Abdullayev, N. A.; Zverev, V. N.L; Alfonsovet, A. et al. Prediction and Observation of an Antiferromagnetic Topological Insulator. *Nature* **2019**, 576, 416-422.

28. Zhang, D. Q.; Shi, M. J.; Zhu, T. S.; Xing, D. Y.; Zhang, H. J. & Wang, J. Topological Axion States in the Magnetic Insulator $MnBi_2Te_4$ with the Quantized Magnetoelectric Effect. *Phys. Rev. Lett.* **2019**, 122, 206401.

29. Deng, Y. J.; Yu, Y. J.; Shi, M. Z.; Guo, Z. G.; Xu, Z. H.; Wang, J.; Chen, X. H. & Zhang, Y. B. Quantum Anomalous Hall Effect in Intrinsic Magnetic Topological Insulator $MnBi_2Te_4$. *Science* **2020**, 367, 895-900.

30. Liu, C.; Wang, Y. C.; Li, H.; Wu, Y.; Li, Y. X.; Li, J. H.; He, K.; Xu, Y.; Zhang, S. S. & Wang, Y. Y. Robust Axion Insulator and Chern Insulator Phases in a Two-Dimensional Antiferromagnetic Topological Insulator. *Nat. Mater.* **2020**, 19, 522-527.

31. Gu, M.; Li, J.; Sun, H.; Zhao, Y.; Liu, C.; Liu, J.; Lu, H.; Liu, Q. Spectral Signatures of the Surface Anomalous Hall Effect in Magnetic Axion Insulators. *Nat. Commun.* **2021**, 12, 3524.

32. Hao, Y. J.; Liu, P. F.; Feng, Y.; Ma, X. M.; Schwier, E. F.; Arita, M.; Kumar, S.; Hu, C. W.; Lu, R.; Zeng, M.; Wang, Y.; Hao, Z. Y.; Sun, H. Y.; Zhang, K.; Mei, J. W.; Ni, N.; Wu, L. S.; Shimada, K. Y.; Chen, C. Y.; Liu, Q. H.; et al. Gapless Surface Dirac Cone in Antiferromagnetic Topological Insulator $MnBi_2Te_4$. *Phys.*

*Rev. X* **2019**, 9, 041038.

33. Li, H.; Gao, S. Y.; Duan, S. F.; Xu, Y. F.; Zhu, K. J.; Tian, S. J.; Gao, J. C.; Fan, W .H.; Rao, Z. C.; Huang, J. R.; Li, J. J.; Yan, D. Y.; Liu, Z. T.; Liu, W. L.; Huang, Y. B.; et al. Dirac Surface States in Intrinsic Magnetic Topological Insulators $EuSn_2As_2$ and $MnBi_{2n}Te_{3n+1}$. *Phys. Rev. X* **2019**, 9, 041039.

34. Chen, Y. H.; Xu, L. X.; Li, J. H.; Li, Y. W.; Wang, H. Y.; Zhang, C. F.; Li, H.; Wu, Y.; Liang, A. J.; Chen, C.; Jung, S. W.; Cacho, C.; Mao, Y. H.; Liu, S.; Wang, M. X.; Guo, Y. F.; Xu, Y.; Liu, Z. K.; Yang, L. X. & Chen, Y. L. Topological Electronic Structure and Its Temperature Evolution in Antiferromagnetic Topological Insulator $MnBi_2Te_4$, *Phys. Rev. X* **2019**, 9, 041040.

35. Wu, J. Z.; Liu, F. C.; Sasase, M.; Ienage, K.; Obata, Y.; Yukawa, Y.; Horiba, K.; Kumigashira, K.; Okuma, S.; Inoshita, T. & Hosono, H. Natural Van Der Waals Heterostructural Single Crystals with both Magnetic and Topological Properties. *Sci. Adv* **2019**, 5, eaax9989.

36. Sun, H. Y.; Xia, B. W.; Chen, Z. J.; Zhang, Y. J.; Liu, P. F.; Yao, Q. S.; Tang, H.; Zhan, Y. J.; Xu, H. & Liu, Q. H. Rational Design Principles of the Quantum Anomalous Hall Effect in Superlatticelike Magnetic Topological Insulators. *Phy. Rev. Lett.* **2019**, 123, 096401.

37. Ge, J.; Liu, Y. Z.; Li, J. H.; Li, H.; Luo, T. C.; Wu, Y.; Xu, Y. & Wang, J. High-Chern-Number and High-Temperature Quantum Hall Effect without Landau Levels. *National Science Review* **2020**, 7, nwaa089.

38. Ovchinnikov, D.; Huang, X.; Lin, Z.; Fei, Z.; Cai, J.; Song, T.; He, M.; Jiang, Q.;

Wang, C.; Li, H.; Wang, Y.; Wu, Y.; Xiao, D.; Chu, J. H.; Yan, J.; Chang, C. Z.; Cui, Y. T.; Xu, X. Intertwined Topological and Magnetic Orders in Atomically Thin Chern Insulator MnBi$_2$Te$_4$. *Nano Lett.* **2021**, 21, 2544-2550.

39. Yang, S.; Xu, X.; Zhu, Y.; Niu, R.; Xu, C.; Peng, Y.; Cheng, X.; Jia, X.; Huang, Y.; Xu, X.; Lu, J.; Ye, Y. Odd-Even Layer-Number Effect and Layer-Dependent Magnetic Phase Diagrams in MnBi$_2$Te$_4$. *Phys. Rev. X* **2021**, 11, 011003.

40. Zhang, R. X.; Wu, F. & Sarma, S. D. Möbius Insulator and Higher-Order Topology in MnBi$_{2n}$Te$_{3n+1}$. *Phys. Rev. Lett.* **2020**, 124, 136407.

41. Xie, H. K.; Wang, D. H.; Cai, Z. X.; Chen, B.; Gou, J. W.; Naveed, M.; Zhang, S.; Zhang, M. H.; Wang, X. F.; Fei, F. C.; Zhang, H. J. & Song, F. Q. The Mechanism Exploration for Zero-Field Ferromagnetism in Intrinsic Topological Insulator MnBi$_2$Te$_4$ by Bi$_2$Te$_3$ Intercalations. *Appl. Phys. Lett.* **2020**, 116, 221902.

42. Hui, C. W.; Gordon, K. N.; Liu, P. F.; Liu, J. Y.; Zhao, X. Q.; Hao, P. P.; Narayan, D.; Emmanouilidou, E.; Sun, H. Y.; Liu, Y. T.; Brawer, H.; Ramirez, A. P.; Ding, L.; Cao, H. B.; Liu, Q. H.; Dessau, D. & Ni, N. A Van Der Waals Antiferromagnetic Topological Insulator with Weak Interlayer Magnetic Coupling. *Nat. Commun.* **2020**, 11, 97.

43. Nevola, D.; Li, H. X.; Yan, J. Q.; Moore, R. G.; Lee, H. N.; Miao, H. & Johnson, P. D. Coexistence of Surface Ferromagnetism and a Gapless Topological State in MnBi$_2$Te$_4$, *Phys. Rev. Lett.* **2020**, 125, 117205.

44. Yuan, Y. H.; Wang, X. T.; Li, H.; Li, J. H.; Ji, Y.; Hao, Z. Q.; Wu, Y.; He, K.; Wang, Y. Y.; Xu, Y.; Duan, W. H.; Li, W. & Xue, Q. K. Electronic States and Magnetic Response of MnBi$_2$Te$_4$ by Scanning Tunneling Microscopy and Spectroscopy. *Nano. Lett.* **2020**, 20, 3271-3277.

45. Wu, X. F.; Li, J. Y.; Ma, X. M.; Zhang, Y.; Liu, Y. T.; Zhou, C. S.; Shao, J. F.; Wang, Q. M.; Hao, Y. J.; Feng, Y.; Schwier, E. F.; Kumar, S.; Sun, H. Y.; Liu, P. F.; Shimada, K.; Miyamoto, K.; Okuda, T.; Wang, K. D.; Xie, M. H.; Chen, C. Y.; Liu, Q. H.; Liu, C. & Zhao, Y. Distinct Topological Surface States on the Two Terminations of MnBi$_4$Te$_7$. *Phy. Rev. X* **2020**, 10, 031013.

46. Liang, Z. W.; Luo, A. Y.; Shi, M. Z.; Zhang, Q.; Nie, S. M.; Ying, J. J.; He, J. F.; Wu, T.; Wang, Z. J.; Xu, G.; Wang, Z. Y. & Chen, X. H. Mapping Dirac Fermions in the Intrinsic Antiferromagnetic Topological Insulators (MnBi$_2$Te$_4$)(Bi$_2$Te$_3$)$_n$ (n=0,1). *Phy. Rev. B* **2020**, 102, 161115.

47. Ko, W.; Kolmer, M.; Yan, J. Q.; Pham, A. D.; Fu, M. M.; Lüpke, F.; Okamoto, S.; Gai, Z.; Ganesh, P. & Li, A. P. Realizing Gapped Surface States in the Magnetic Topological Insulator MnBi$_{2-x}$Te$_4$. *Phys. Rev. B* **2020**, 102, 115402.

48. Sass, P. M.; Kim, J. W.; Vanderbilt, D.; Yan, J. Q. & Wu, W. D. Robust A-type Order and Spin-Flop Transition on the Surface of the Antiferromagnetic Topological Insulator MnBi$_2$Te$_4$. *Phy. Rev. Lett.* **2020**, 125, 037201.

49. Jia, B.; Zhang, S.; Ying, Z.; Xie, H. K.; Chen, B.; Naveed, M.; Fei, F. C. & Zhang, M. H. Unconventional Anomalous Hall Effect in Magnetic Topological Insulator MnBi$_4$Te$_7$ device. *Appl. Phys. Lett.* **2021**, 118. 083101.

50. Lu, R.; Sun, H. Y.; Kumar, S.; Wang, Y.; Gu, M. Q.; Zeng, M.; Hao, Y. J.; Li, J. Y.; Shao, J. F., Ma, X. M.; Hao, Z. Y.; Zhang, K.; Mansuer, W.; Mei, J. W.; Zhao, Y.;

Liu, C.; Deng, E.; Huang, W.; Shen, B.; Shimada, K. et al. Half-Magnetic Topological Insulator with Magnetization-Induced Dirac Gap at a Selected Surface. *Phys. Rev. X* **2021**, 11, 011039.

51. Chen, W. Z.; Zhao, Y. F.; Yao, Q. S.; Zhang, J. & Liu, Q. H.; Koopmans' Theorem as the Mechanism of Nearly Gapless Surface States in Self-Doped Magnetic Topological Insulators. *Phys. Rev. B* **2021**, 103, L201102.

52. Ma, X. M. et al. Realization of a Tunable Surface Dirac Gap in Sb-Doped $MnBi_2Te_4$. *Phys. Rev. B* **2021**, 103, L121112.

53. Sekine, A. & Nomura, K. Axion Electrodynamics in Topological Materials. *J. Appl. Phys.* **2021**, 129, 141101.

54. Zhang, J. L.; Wang, D. H.; Shi, M. J.; Zhu, T. S.; Zhang, H. J.; Wang, J. Large Dynamical Axion Field in Topological Antiferromagnetic Insulator $Mn_2Bi_2Te_5$, *Chin. Phys. Lett.* **2020**, 37, 077304.

55. Jo, N. H.; Wang, L. L.; Slager, R. J.; Yan, J. Q.; Wu, Y.; Lee, K.; Schrunk, B.; Vishwanath, A. & Kaminski, A. Intrinsic Axion Insulating Behavior in Antiferromagnetic $MnBi_6Te_{10}$. *Phys. Rev. B* **2020**, 102, 045130.

56. Rienks, E. D. L.; Wimmer, S.; Sánchez-Barriga, J.; Caha, O.; Mandal, P. S.; Růžička, J.; Ney, A.; Steiner, H.; Volobuev, V. V.; Groiss, H.; Albu, M.; Kothleitner, G.; Michalička, J.; Khan, S. A.; Minár, J.; Ebert, H.; Bauer, G.; Freyse, F.; Varykhalov, A.; Rader, O. et al. Large Magnetic Gap at the Dirac Point in $Bi_2Te_3$/$MnBi_2Te_4$ Heterostructures. *Nature* **2019**, 576, 423–428.

57. Zhu, T. S.; Wang, H. Q.; Zhang, H. J. & Xing, D. Y. Tunable Dynamical

Magnetoelectric Effect in Antiferromagnetic Topological Insulator MnBi$_2$Te$_4$ Films. *npj Comput. Mater.* **2021**, 7, 121

58. Hu, C. W.; Ding, L.; Gordon, K. N.; Ghosh, B.; Tien, H. J.; Li, H. X.; Linn, A. G.; Lien, S. W.; Huang, C. Y.; Mackey, S.; Liu, J. Y.; Reddy, P. V. S.; Singh, B.; Agarwal, A.; Bansil, A.; Song, M.; Li, D. S.; Xu, S. Y.; Lin, H.; Cao, H. B. et al. Realization of an Intrinsic Ferromagnetic Topological State in MnBi$_8$Te$_{13}$. *Sci. Adv.* **2020**, 6, eaba4275.

59. Alpichshev, Z.; Biswas, R. R.; Balatsky, A. V.; Analytis, J. G.; Chu, J. H.; Fisher, I. R. & Kapitulnik, A. STM Imaging of Impurity Resonances on Bi$_2$Se$_3$. *Phys. Rev. Lett.* **2012**, 108, 206402.

60. Varnava, N. & Vanderbilt, D. Surfaces of Axion Insulators. *Phy. Rev. B* **2018**, 98, 245117.

61. Lüpke, F.; Pham, A. D.; Zhao, Y. F.; Zhou, L. J.; Lu, W.; Briggs, E.; Bernholc, J.; Kolmer, M.; Ko, M.; Chang, C. Z.; Ganesh, P. & Li, A. P. Local Manifestations of Thickness Dependent Topology and Axion Edge State in Topological Magnet MnBi$_2$Te$_4$. *Phys. Rev. B* **2022,** 105, 035423.

62. Lin, W.; Feng, Y.; Wang, Y.; Lian, Z.; Li, H.; Wu, Y.; Liu, C.; Wang, Y.; Zhang, J.; Wang, Y.; Zhou X. & Shen, J. Direct Visualization of Edge State in Even-Layer MnBi$_2$Te$_4$ at Zero Magnetic Field. **2021,** 2105.10234, ArXiv[cond-mat], https://doi.org/10.48550/arXiv.2105.10234 (accessed May 21, 2021)

63. Blöchl, P. E. Projector Augmented-Wave Method. *Phys. Rev. B* **1994**, 50, 17953.

64. Ropo, M.; Kokko, K. & Vitos, L. Assessing the Perdew-Burke-Ernzerhof

Exchange-Correlation Density Functional Revised for Metallic Bulk and Surface Systems. *Phys. Rev. B* **2008**, 77, 195445.

65. Kresse, G. & Furthmüller, J. Efficiency of Ab-Initio Total Energy Calculations for Metals and Semiconductors using a Plane-Wave Basis Set. *Computational Materials Science* **1996**, 6, 15-50.

66. Kresse, G. & Joubert, D. From Ultrasoft Pseudopotentials to the Projector Augmented-Wave Method. *Phys. Rev. B* **1999**, 59, 1758.

67. Dudarev, S. L.; Botton, G. A.; Savrasov, S. Y.; Humphreys, C. J. & Sutton, A. P. Electron-Energy-Loss Spectra and the Structural Stability of Nickel Oxide: An LSDA+U study. *Phys. Rev. B* **1998**, 57, 1505.

68. Wu, Q.; Zhang, S.; Song, H. F.; Troyer, M. & Soluyanov, A. A. WannierTools: An Open-Source Software Package for Novel Topological Materials. *Comput. Phys. Commun.* **2018**, 224, 405.

69. Mostofi, A. A. Wannier90: A Tool for Obtaining Maximally-Localised Wannier Functions. *Comput. Phys. Commun.* **2008**, 178, 685.

70. Marzari, N.; Mostofi, A. A.; Yates, Y. R.; Souza, I. & Vanderbilt, D.; Maximally Localized Wannier Functions: Theory and Applications. *Rev. Mod. Phys.* **2012**, 84, 1419.